\documentclass[aps,twocolumn,showpacs,showkeys,preprintnumbers,amsmath,amssymb]{revtex4}

\usepackage{epsfig}

\def\be{\begin{equation}}

\def\ee{\end{equation}}

\def\lsim{\raise0.3ex\hbox{$<$\kern-0.75em\raise-1.1ex\hbox{$\sim$}}}

\def\gsim{\raise0.3ex\hbox{$>$\kern-0.75em\raise-1.1ex\hbox{$\sim$}}}

\begin{document}

\title{Critical Droplets and Phase Transitions in Two Dimensions}

\vskip 1.cm

\author{Santo Fortunato}

\vskip0.7cm

\affiliation{Fakult\"at f\"ur Physik, Universit\"at Bielefeld, D-33501 Bielefeld, Germany}

\vskip0.5cm

\begin{abstract}
\noindent

In two space dimensions, the percolation 
point of the pure-site clusters of the Ising model coincides with the
critical point $T_c$ of the thermal transition and
the percolation exponents belong to
a special universality class. By introducing a bond probability
$p_B<1$, the corresponding site-bond clusters keep on percolating
at $T_c$ and the exponents do not change, until $p_B=p_{CK}=1-\exp(-2J/kT)$:
for this special expression of the bond weight the critical percolation exponents
switch to the 2D Ising universality class.
We show here that
the result is valid for a wide class of bidimensional 
models with a continuous magnetization
transition: there is a critical bond probability $p_c$ such that,
for any $p_B\,{\geq}\,p_c$, the onset of percolation of the site-bond clusters 
coincides with the critical point of the thermal transition. The percolation 
exponents are the same for $p_c<p_B\,{\leq}\,1$ but, for $p_B=p_c$, 
they suddenly change to the thermal exponents, so that the 
corresponding clusters are critical droplets of the phase
transition. Our result is 
based on Monte Carlo simulations of various systems near criticality.

\end{abstract}



\maketitle

\vskip0.7cm

\section{Introduction}

The purely geometric percolation phenomenon \cite{stauffer} has astonishing
analogies with ordinary second-order thermal phase transitions:
the percolation variables show power-law behaviour near criticality,
with relative exponents; the exponents are not all independent, but 
related to each other by scaling equalities which are formally
identical to the corresponding thermal ones; the critical indices 
do not depend on the specific features
of a system but they
are grouped in universality classes, which in the percolation case
are uniquely specified by the number of space dimensions
of the system.

These analogies suggested many years ago that
a continuous thermal transition might be 
nothing but a percolation transition \cite{fisher},
provided one defines suitable clusters, or {\it droplets}.
The growth of the droplets describes naturally the 
propagation of correlations between different particles of the system 
and the formation of an infinite spanning structure 
represents the long-range order of the system in
the new phase. 

In this paper we focus on spin models with a continuous magnetization transition.
In order to map the critical behaviour of a system onto a percolation
picture, one must set a correspondence between thermal and geometric variables.
The main percolation 
variables are:

\begin{itemize}
\item{the percolation strength $P$, i.e. the probability that a site chosen
at random belongs to a percolating cluster;}
\item{the average cluster size $S$,
\begin{equation}\label{defS}
  S\,=\,\frac{\sum_{s} {{n_{s}s^2}}}{\sum_{s}{n_{s}s}}~,
\end{equation}
where $n_s$ is the number of clusters with $s$ sites and the sums exclude eventual
percolating clusters.
}
\end{itemize}
The conditions for a successful 
mapping are listed below:
\begin{itemize}
\item{the percolation point must coincide with the thermal critical point;}
\item{the connectedness length (average cluster radius) diverges as the 
thermal correlation length (same exponent);}
\item{the percolation strength $P$ near the threshold
varies like the order parameter $m$
of the model (same exponent);}
\item{the average cluster size $S$ diverges as the 
physical susceptibility $\chi$ (same exponent).}
\end{itemize}
It is not yet known whether, given a system, 
there are at all clusters satisfying such conditions, which are quite strict. 
The first studies concentrated on the
simplest theory, the Ising model without field. A real breakthrough in these
investigations was the discovery that the Ising model can be rewritten as a 
geometrical model \cite{fortuin}, where the fundamental objects are
special site-bond clusters, i.e. clusters built by joining 
nearest-neighbouring spins of the same sign with a temperature-dependent
bond probability $p_{CK}=1-\exp(-2J/kT)$ ($J$ is the Ising spin-spin coupling).
It is possible to prove that these site-bond clusters are just the critical
droplets of the system. Starting from this result, which is valid
more in general for the $q$-state Potts model, it is possible to define
the droplets for a wide variety of systems, like models with
several spin-spin couplings, as long as they are all ferromagnetic
\cite{san}, and $O(n)$ spin models \cite{san1}.
A common feature of all these results is the fact that a geometrical bond
(with a bond probability) is associated with each spin-spin coupling 
in the percolation picture. 
In this way, for models with several (ferromagnetic) interactions,
the clusters are quite weird objects, consisting of 
geometrically disconnected
parts which are joined to each other by virtue of invisible 
bridges due to long-ranged interactions. 
Besides, if not all couplings are ferromagnetic, it is not yet clear
if and how one can define the droplets.

Very recent results show that the simple connections 
between nearest-neighbour spins
have a close relationship with the phase transition of several bidimensional
models \cite{santo}: the percolation point of the pure-site clusters 
coincides with the critical point and the percolation exponents,
which differ from the thermal exponents, are the same for models
in the same universality class. In our opinion this 
property indicates that, in many cases,
the simple geometrical connectivity between nearest-neighbours
plays a crucial role in the mechanism
of the phase transition.

In the 2D Ising model, the fact that both 
the pure-site and the site-bond clusters with bond probability
$p_{CK}$ start
to percolate at the critical
temperature $T_c$ necessarily implies that, by taking 
a bond probability $p_B$ such that $p_{CK}<p_B\,{\leq}\,1$,
the percolation temperature of the site-bond
clusters with bond weight $p_B$ is always $T_c$. A renormalization group
analysis \cite{CK} led to the conclusion that the percolation exponents
are the same as for the pure-site clusters for any $p_B>p_{CK}$, and that they 
suddenly switch to the thermal exponents for $p_B=p_{CK}$. For
$p_B<p_{CK}$, the site-bond clusters percolate at some 
temperature $T_p<T_c$ and the exponents switch to the 2D random percolation
universality class. 
This analysis, though reliable, is however not rigorous. In this paper
we will show that the result is true and valid
for many bidimensional models. We will see that, given a model,
there is a critical probability $p_c$ such that, by taking a bond
weight $p_B>p_c$ (including the pure-site case
$p_B=1$) the site-bond clusters keep percolating at $T_c$
and the critical percolation exponents are the same; for $p_B=p_c$
the exponents change abruptly to the thermal exponents. The probability 
$p_c$, which is model-dependent, 
is the minimal probability which still allows the formation
of a percolating cluster at $T_c$, and this gives us a general
prescription for the critical droplets.

\section{Results of the simulations}

Our investigations consisted of Monte Carlo simulations on square
lattices of 
several models around the critical temperature. We took the same models
that were studied in \cite{santo}, i.e.

\begin{enumerate}
\item{the Ising model, ${\cal{H}}=-J\sum_{ij}s_is_j$ \,\,\,($J>0$, $s_i={\pm}1$);}
\item{a model with nearest-neighbour (NN) ferromagnetic coupling and 
a weaker next-to-nearest (NTN) antiferromagnetic coupling: 
${\cal{H}}=-J_1\sum_{NN}s_is_j-J_2\sum_{NTN}s_is_j$ ($J_1>0$,
$J_2<0$, $|J_2/J_1|=1/10$, $s_i={\pm}1$);}
\item{the continuous Ising model, ${\cal{H}}=-J\sum_{ij}S_iS_j$ 
\,\,\,($J>0$, $-1{\leq}S_i{\leq}+1$);}
\item{SU(2) pure gauge theory in 2+1 dimensions;}
\item{the 3-state Potts model, ${\cal{H}}=-J\sum_{ij}\delta(s_i,s_j)$ \,\,\,($J>0$, $s_i=1,2,3$);}
\item{a model obtained by adding to 5) 
a weaker next-to-nearest (NTN) antiferromagnetic coupling: 
${\cal{H}}=-J_1\sum_{NN}\delta(s_i,s_j)-J_2\sum_{NTN}\delta(s_i,s_j)$ \,\,\,($J_1>0$,
$J_2<0$, $|J_2/J_1|=1/10$, $s_i=1,2,3$);}
\end{enumerate}
The critical temperatures
of all these systems are known with good precision (see \cite{santo}), for Models 1 and 5 
the values are analytically known. We stress that the models belong to the 2D
$Z(2)$ (1-4) and $Z(3)$ (5,6)
universality classes. 
The estimates of \cite{santo} for  
the main critical indices of the percolation transition
of the pure-site clusters for the two classes are in very good agreement with
theoretical predictions \cite{CK}-\cite{conpe}; we list
these analytical values in Table \ref{tab1}.

\begin{table}[htb]
\begin{center}
\begin{tabular}{|c|c|c|c|c|c|}
\hline$\vphantom{\displaystyle\frac{1}{1}}$
&$\beta_p$ &$\gamma_p$  & $\nu_p$ & Fractal Dim. D & Cum. at $T_p$\\
\hline$\vphantom{\displaystyle\frac{1}{1}}$
Z(2) & 5/96 & 91/48 & 1 & 187/96 &0.9832(4)\\
\hline$\vphantom{\displaystyle\frac{1}{1}}$
Z(3) &  7/96 & 73/48 & 5/6 & 153/80 &0.932(2)\\
\hline
\end{tabular}
\caption{\label{tab1} Critical percolation indices for pure-site
clusters for the two universality classes we considered.}
\end{center}
\end{table}

For the simulations we made use of standard algorithms like Metropolis or heat
bath; in some cases we could apply the Wolff cluster algorithm, which 
allowed us to reduce sensibly the correlation of the data.
For each system we took four to six different lattice sizes.
The clusters were identified by means of the algorithm 
devised by Hoshen and Kopelman \cite{kopelman}. 
We used everywhere free boundary 
conditions for the cluster labeling and
say that a cluster percolates if it connects the top with the bottom side
of the lattice. At each iteration we calculated the variables
$P$ and $S$ and the size $S_M$ of the
largest cluster, from which one can determine
the fractal dimension $D$ of the percolating cluster at $T_c$.
Moreover, from the data sample of the percolation strength $P$
one can extract a variable which turns out to be a formidable tool to investigate
numerically the percolation transition. In fact, the number of
configurations with (at least) a percolating cluster (i.e. for which $P{\neq}0$)
for a given temperature and
lattice size, divided by the total number of configurations, returns
a scaling variable $\Pi$ called percolation cumulant, whose properties are identical
to those of the Binder cumulant in standard thermal transitions \cite{binder}. 
In particular, the value of $\Pi$ at the critical threshold
is a universal quantity, i.e. it is the same for models in the same universality class. 
We determined the percolation exponents through
standard finite size scaling tecniques at the critical temperature 
$T_p$, considering
simply the leading behaviour
 
\begin{eqnarray}
P(T_p)\,&\propto&\,L^{-\beta_p/\nu_p}\\
S(T_p)\,&\propto&\,L^{\gamma_p/\nu_p}\\
S_M(T_p)\,&\propto&\,L^{D}.
\end{eqnarray}

\begin{figure}[htb]
  \begin{center}
    \epsfig{file=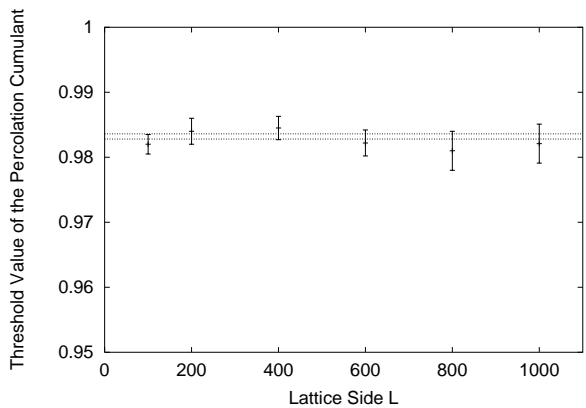,width=9cm}
    \caption{\label{fig1}{2D Ising model: value of the percolation
        cumulant $\Pi$ at the critical temperature 
        $T_c$ for various lattices. The bond weight 
        is $CK_2=1-\exp(-2.5J/kT)$. All values agree with each other (so $T_p=T_c$)
        and with the 
        (universal) pure-site value, represented with its error by the dashed
        lines in the plot.}}
  \end{center}
\end{figure}

\begin{figure}[htb]
  \begin{center}
    \epsfig{file=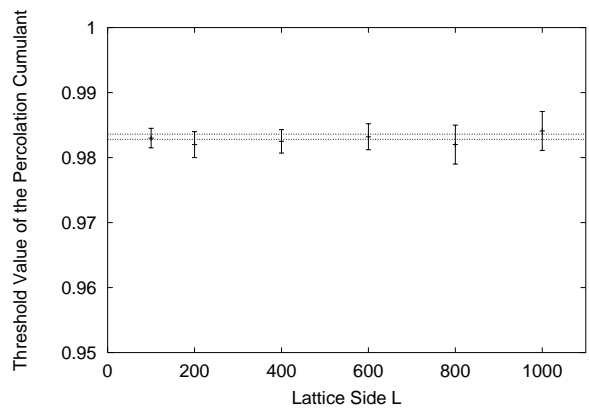,width=9cm}
    \caption{\label{fig2}{2D Ising model: value of the percolation
        cumulant $\Pi$ at 
        the critical temperature $T_c$ for various lattices. The bond weight 
        is $CK_3=1-\exp(-2.2J/kT)$. All values agree with each other (so
        $T_p=T_c$) and with the 
        (universal) pure-site value, represented with its error by the dashed
        lines in the plot.}}
  \end{center}
\end{figure}

We began our studies starting from the Ising model. We tested three different expressions
for the bond probability \cite{nota1}
$p_B$: $CK_1=1-\exp(-3J/kT)$, $CK_2=1-\exp(-2.5J/kT)$, $CK_3=1-\exp(-2.2J/kT)$.
In all cases we found that, even for the smallest lattice size we examined
($100^2$), the threshold value of the percolation cumulant $\Pi$ did not differ from
the value relative to the pure-site clusters ($0.9832(4)$),
which is a clear indication that the critical indices
of the three percolation transitions remain in the
universality class of the 2D $Z(2)$ pure-site clusters. 
Figs. \ref{fig1} and \ref{fig2} show the situation for the
probabilities $CK_2$ and $CK_3$, respectively.

Indeed, the values of the
percolation exponents we extracted
from finite size scaling fits of $P$, $S$ and $S_M$ 
at the critical point are in good agreement with the $Z(2)$ values 
reported in Table \ref{tab1}. Next, we approached the critical value of
$p_{CK}$ at $T_c$ ($p_{CK}(T_c)=0.58578...$) by successive numerical trials.
We found that, up to a value greater than $p_{CK}(T_c)$ by $4-5\%$, the
threshold value of the cumulant did not change for any of the lattices we have
taken, and the situation is the same as
in Figs. \ref{fig1} and \ref{fig2}. 
For still lower probabilities the cumulant decreases appreciably for
the smaller lattices and the less the larger the size. That is 
the "lattice reaction" to the fact that we are near an abrupt discontinuity.
We know that the finite size of
the lattice reduces divergences
to finite peaks and discontinuities to smooth variations.
Close to a sharp discontinuity, the real behaviour of the system 
is reproduced only on very large lattices; for small sizes,
the system feels the proximity 
of the new state
and that gives rise to intermediate configurations, which are unphysical.
On the other hand, the threshold value of the percolation cumulant on our
largest lattice ($1000^2$) remained fixed at the pure-site value until
$p_B/p_{CK}(T_c)\,{\approx}\,1.02$, and that clearly suggests
that the transition from the pure-site behaviour to the droplet behaviour
takes place just when $p_B=p_{CK}(T_c)$.  

\begin{figure}[htb]
  \begin{center}
    \epsfig{file=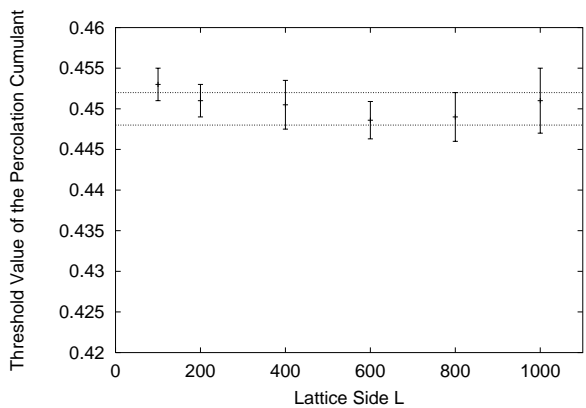,width=9cm}
    \caption{\label{fig31}{2D Ising model: value of 
        $\Pi$ at 
        the percolation temperature $T_p$ for various lattices. The bond weight 
        is $1-\exp(-J/kT)<p_c=1-\exp(-2J/kT)$. In this case 
        is $T_p<T_c$. The data points agree 
        with the (universal) value for 2D random percolation, represented with its error by the dashed
        lines in the plot.}}
  \end{center}
\end{figure}

By taking exactly $p_B=p_{CK}(T_c)$, the threshold value of the percolation cumulant
is again the same for all lattices, and equal to the value $0.585(1)$ that
labels the universality class of the 2D $Z(2)$ droplets. Next, we took some
values of $p_B<p_{CK}(T_c)$. Just below $p_{CK}(T_c)$ we expected
to see the same mixed behaviour observed when one approaches 
the critical probability from above, and that is indeed the case. However,
already for $p_B/p_{CK}(T_c)<0.96$ we could see again a "clean" behaviour
of the percolation variables on all lattices:
the percolation cumulants no longer cross
at $T_c$ but at a somewhat lower temperature $T_p$, which decreases
by decreasing $p_B$.
The value of the cumulant at the crossing point is the same for any $p_B$ and agrees
with the value corresponding to the 2D random percolation universality class
($0.450(2)$), as shown in Fig. \ref{fig31}. 
Our findings confirm the scenario predicted in \cite{CK}.

As far as
the other models with an Ising-like transition are concerned
we remark that, except for the continuous Ising model \cite{san2}, the 
problem of the existence of 
eventual critical droplets is open. Since in each case the
pure-site clusters begin to percolate at $T_c$, it was natural
for us to check whether also here one has the same situation like
in the Ising model. 
We then sat at $T_c$ in every case, and introduced 
progressively decreasing bond weights. We remark that
for the continuous spin models 3 and 4 we bound nearest-neighbour
spins of the same sign, independently of their absolute value. 

\begin{table}[h]
\begin{center}
\begin{tabular}{|c|c|c|c|c|}
\hline$\vphantom{\displaystyle\frac{1}{1}}$
&$\beta_p/\nu_p$ &$\gamma_p/\nu_p$  & Fractal Dim. D & Cum. at $T_p$\\
\hline$\vphantom{\displaystyle\frac{1}{1}}$
2D Ising & 1/8=0.125 & 7/4=1.75&15/8=1.875 &0.585(1)\\
\hline$\vphantom{\displaystyle\frac{1}{1}}$
Model 2 & 0.131(10) & 1.742(12)&1.862(20)&0.583(4)\\
\hline$\vphantom{\displaystyle\frac{1}{1}}$
Model 3 & 0.121(9) & 1.764(14)&1.870(11)&0.587(3)\\
\hline$\vphantom{\displaystyle\frac{1}{1}}$
Model 4 & 0.140(19) & 1.761(17)&1.882(18)&0.586(5)\\
\hline
\end{tabular}
\caption{\label{tab2} Critical percolation indices for the site-bond 
clusters of the models with an
  Ising-like transition when $p_B=p_c$, compared with the values of the
2D Ising droplets.}
\end{center}
\end{table}

\begin{figure}[htb]
  \begin{center}
    \epsfig{file=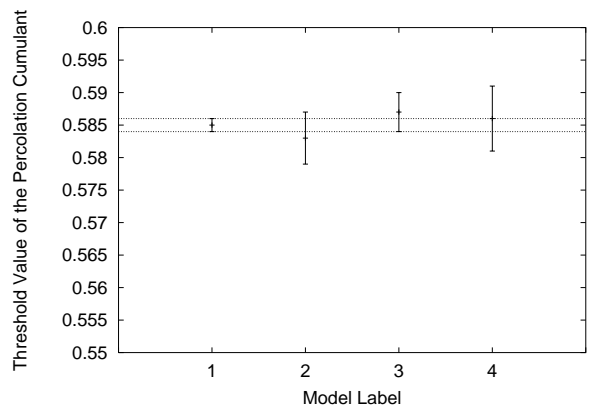,width=9cm}
    \caption{\label{fig3}{Threshold value of the percolation
        cumulant $\Pi$ at 
the critical temperature $T_c$ by using the "minimal" probability
$p_c$ for all $Z(2)$ models. The $X$-coordinates 1, 2, 3 and 4 label
the four models we investigated.  
There is a clear accord of the four data points with each other.
The dashed lines indicate the bounds of the more precisely 
measured Ising value (first point to the left), 
which labels the universality class of the 2D Ising  
droplets.}}
\end{center}
\end{figure}

We found that
the scenario is indeed the same. In every case, we found a critical
probability $p_c$, which equals 0.583(1), 0.6115(9) and 0.6275(7) for 
Model 2,3 and 4, respectively. We adopted several values for $p_B$,
both above and below $p_c$. For $p_B>p_c$ we have always found that
the percolation point coincides with the critical temperature $T_c$ of the
model, and that the critical percolation indices coincide (within errors) with the
$Z(2)$ percolation indices of Table \ref{tab1}. In all cases we could
show analogous pictures as Fig. \ref{fig1} or \ref{fig2}. If $p_B<p_c$, the 
percolation temperature $T_p<T_c$ and the exponents belong to the
2D random percolation universality class (as in Fig. \ref{fig31}). Like in the Ising
model $p_c$ is then the
minimal probability for which the site-bond clusters still percolate at $T_c$.
Fig. \ref{fig3} shows the threshold value of the 
percolation cumulant for all $Z(2)$ models when $p_B=p_c$; all values agree
within errors. In this case, the critical indices are in accord with the ones of the 2D Ising
droplets,
as shown in Table \ref{tab2}. 

\begin{figure}[htb]
  \begin{center}
    \epsfig{file=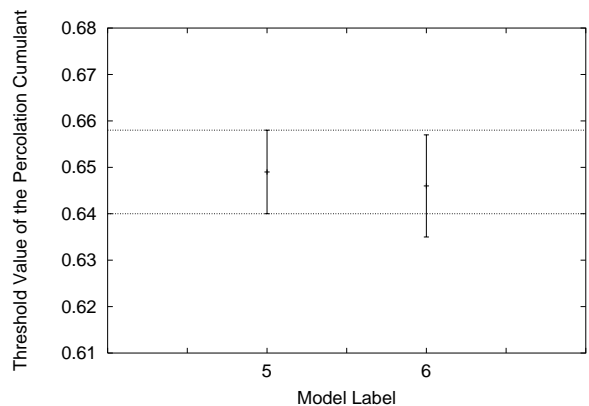,width=9cm}
    \caption{\label{fig5}{Threshold value of the percolation
        cumulant $\Pi$ at 
the critical temperature $T_c$ by using the "minimal" probability
$p_c$ for the $Z(3)$ models. The $X$-coordinates 5 and 6 label
the two models we investigated. The values agree well within errors.
The dashed lines indicate the bounds of the more precisely 
measured 3-state Potts value (point to the left), 
which labels the universality class of the 2D $Z(3)$ droplets.}}
\end{center}
\end{figure}

Finally we investigated Models 5 and 6. 
We know that the $q$-state Potts model can be transformed into a cluster
model by means of the Fortuin-Kasteleyn transformation \cite{fortuin}. The
clusters are defined in the same way as for Ising, except that 
the bond weight $p_{FK}=1-\exp(-J/kT)$, but the situation is identical.
It is then not surprising that for the 2D $q$-state Potts
model \cite{nota2} the same scenario as for Ising was predicted \cite{conpe}.
The results of our simulations confirm such prediction. The situation does not
change for Model 6: we found a critical bond probability $p_c=0.61(1)$, at which
the site-bond clusters become critical droplets for the model, as one
can see in Table \ref{tab3}. In Fig. \ref{fig5} we compare the 
threshold values of $\Pi$ for the two $Z(3)$ models when $p_B=p_c$.

\begin{table}[h]
\begin{center}
\begin{tabular}{|c|c|c|c|c|}
\hline$\vphantom{\displaystyle\frac{1}{1}}$
&$\beta_p/\nu_p$ &$\gamma_p/\nu_p$  & D & Cum. at $T_p$\\
\hline
\hline$\vphantom{\displaystyle\frac{1}{1}}$
2D 3S Potts & 2/15 & 26/15&28/15&0.649(9)\\
\hline$\vphantom{\displaystyle\frac{1}{1}}$
Model 6 & 0.143(17) & 1.725(21)&1.858(18)&0.646(11)\\
\hline
\end{tabular}
\caption{\label{tab3} Critical percolation indices for the site-bond clusters
of Model 6
when $p_B=p_c$, compared with the corresponding values of the
2D 3-state Potts droplets.}
\end{center}
\end{table}

\section{Conclusions}

We found a general criterion to identify 
the critical droplets of various bidimensional systems. The droplets are the clusters
obtained by joining nearest-neighbour spins of the same sign (-state for Potts-like systems)
with the minimal bond probability $p_c$ that still allows the clusters
to percolate at the critical point of the thermal transition. For any $p_B>p_c$
the onset of 
percolation for the relative site-bond clusters keeps coinciding
with the thermal threshold and the critical indices are the
same up to the pure-site case $p_B=1$. We stress that this prescription seems to work
also for the cases in which one can rigorously define the droplets. For the
continuous Ising model, for instance, the droplets are 
defined by introducing local
bond weights which also depend on the length of the spins and not only on their
sign \cite{san2}. These clusters certainly differ in detail from the droplets defined
in this paper; nevertheless their behaviour at criticality is identical.
The matter would be even more involved for models with several 
ferromagnetic interactions, for which a rigorous 
definition of droplets requires the presence
of longer-range connections than just between nearest-neighbour spins \cite{san}.
The fact that the critical behaviour of the system is reproduced
by simple site-bond clusters, 
no matter how complicated the model is, suggests that the long-range 
fluctuations of the system, which are responsible for the phase transition, 
are embodied in the simple magnetic domains of the system; the bond probability
$p_c$ is necessary in order to destroy additional spin correlations due to 
purely geometrical effects, as already remarked in \cite{CK}. 
Besides, we have
seen that the result is also valid in cases where 
a rigorous definition of droplets is, at present, missing (Models 2, 4 and 6). 
In particular, it remains true for $SU(2)$ pure gauge theory, which is 
a very complicated model involving many different interactions, like 
multispin couplings, long-ranged couplings (ferromagnetic and antiferromagnetic) 
and self-interactions. This shows that the result has some
generality, and it would be interesting to check 
to which extent it is true for 
bidimensional models with a continuous magnetization transition. 
It would be also interesting to check whether simple site-bond 
clusters play a role in the description
of critical behaviour at higher dimensions as well.  

\begin{acknowledgments}

I would like to thank H. Satz, Ph. Blanchard and A. Coniglio 
for many helpful
discussions. I gratefully acknowledge the financial support of
the TMR network ERBFMRX-CT-970122 and the DFG Forschergruppe FOR
339/1-2.

\end{acknowledgments}


\begin{thebibliography}{99}
\bibitem{nota1} We remark that, for
our purposes, the bond probability needs not be a function of the temperature:
in fact, since we are interested in the behaviour at the transition, the only important
thing is the threshold value
of $p_B$, compared with $p_{CK}(T_c)$.
\bibitem{nota2} This is obviously valid just for the cases in which the system undergoes
a continuous magnetization transition, i.e. for q=2,3,4.
\end{thebibliography}
\end{document}